\documentclass[aps,pra,reprint,groupedaddress,amsmath,amssymb,longbibliography]{revtex4-1}

\usepackage{CJK}
\usepackage{graphicx}% Include figure files
\usepackage{dcolumn}% Align table columns on decimal point
\usepackage{bm}% bold math
\usepackage[pdfstartview=FitH, 
CJKbookmarks=true, 
bookmarksnumbered=true, 
bookmarksopen=true, 
colorlinks=true, 
pdfborder=0 0 1, 
citecolor=blue, 
linkcolor=blue, 
linktocpage=true] {hyperref}
\usepackage{epstopdf} % ensure the PDFlatex can work normally.
\begin{document}
% =======================================Head of Paper=======================================================

\title{Hybrid atom-molecule quantum walks in one-dimensional optical lattice}
\author{Ling Lin $^{1,2}$}
\author{Yongguan Ke $^{1,2}$}

\author{Chunshan He $^{3}$}
\altaffiliation{Email: stshcs@mail.sysu.edu.cn}

\author{Chaohong Lee $^{1,2}$}
\altaffiliation{Email: lichaoh2@mail.sysu.edu.cn; chleecn@gmail.com}

\affiliation{$^{1}$Laboratory of Quantum Engineering and Quantum Metrology, School of Physics and Astronomy, Sun Yat-Sen University (Zhuhai Campus), Zhuhai 519082, China}
\affiliation{$^{2}$Key Laboratory of Optoelectronic Materials and Technologies, Sun Yat-Sen University (Guangzhou Campus), Guangzhou 510275, China}
\affiliation{$^{3}$School of Physics, Sun Yat-Sen University (Guangzhou Campus), Guangzhou 510275, China}

\date{\today}

\begin{abstract}
  We study hybrid atom-molecule quantum walks in one-dimensional optical lattices with two interacting bosonic atoms which may be converted into a molecule.
  The hybrid atom-molecule energy bands include a continuum band and two isolated bands, which respectively correspond to scattering states and dressed bound states (DBS's).
  Because of the atom-molecule coupling, the DBS's may appear even in the absence of atom-atom interaction.
  From an initial state of two atoms occupying the same site, in addition to independent quantum walks which correspond to scattering states, correlated quantum walks appear as a signature of DBS's.
  Even if the atom-atom interaction and the atom-molecule coupling are much stronger than the tunneling strengths, independent quantum walks may still appear under certain resonant conditions.
  The correlated quantum walks show two light-cones with different propagation velocities, which can be analytically explained by the effective tunneling strengths of the two different DBS's.
  %
  % Revised on 2018/06/20
  Furthermore, the effective nearest-neighbor tunneling of DBS's can be suppressed to zero, which can be explained by the destructive interference between the atomic pair and the molecule.
\end{abstract}

\maketitle

\section{Introduction}
% Summarize the quantum walks.
Quantum walks (QWs)~\cite{kempe2003quantum}, a direct result of quantum interference of different paths, have been extensively studied in both theory and experiments~\cite{PhysRevA.48.1687, venegas2012quantum, du2003experimental, PhysRevA.65.032310}.
QWs can be exploited to various fields, from universal quantum computing~\cite{childs2009universal}, efficient quantum algorithm~\cite{farhi1998quantum,shenvi2003quantum,ambainis2007quantum,childs2004spatial,JPhysA.41075303}, energy transfer~\cite{ChemicalTransfer}, to topological state detection~\cite{kitagawa2012,xiao2017}.
% Introduce the single-particle quantum walks.
Single-particle QWs have already been implemented by various systems including ultracold atoms~\cite{karski2009quantum}, ultracold ions~\cite{zahringer2010realization}, photoic waveguides~\cite{PhysRevLett.100.170506} and atomic spin-impurities~\cite{fukuhara2013quantum} etc.
Moreover, it has also been demonstrated that single-particle QWs can be implemented via classical waves~\cite{PhysRevA.68.020301}.

% Introduce the two-particle quantum walks, from non-intearcting particles to interacting particles.
Beyond single-particle QWs, two-particle QWs have attracted extensively interests in recent years.
The non-classical correlation between non-interacting particles, i.e., the bunching and anti-bunching behavior, are found to depend strongly on the quantum statistical properties \cite{peruzzo2010quantum,sansoni2012two,PhysRevLett.102.253904}.
On the other hand, interaction between particles in a lattice is believed to be beneficial to universal quantum computation \cite{childs2013universal}.
The interacting two-particle QWs have been discussed and implemented \cite{preiss2015strongly,Ahlbrecht2012,PhysRevA.86.011603,PhysRevA.96.043629}.
The interaction is found to strongly affect the spatial correlations \cite{PRA.90.062301}.
Particularly, the repulsively or attractively interacting (quasi-)particles can form a bound pair \cite{winkler2006repulsively,fukuhara2013microscopic}.
Therefore, besides the independent QWs, there is the co-walking of the bound pair\cite{folling2007direct,PRA.90.062301,preiss2015strongly}.
%%

% Introduce the atom-molecule coupling and relevant results.
Although the QWs of interacting particles have been extensively studied, it still remains unclear about the QWs involving atom-molecule coupling.
According to the two-channel theory~\cite{PhysRevA.77.021601,PhysRevA.78.023617, PhysRevA.83.031607, Grupp2007}, a pair of atoms can be converted into a molecule.
For two bosons in optical lattices, due to the atom-molecule coupling, their energy spectrum includes two isolated bands and a continuum one~\cite{PhysRevA.77.021601, PhysRevA.78.023617, PhysRevA.83.031607, Grupp2007}.
The states in isolated bands are in superposition of atomic bound state and molecular state, which are called the dressed bound states (DBS's) in the following context.
Under specific conditions, the DBS's can be tuned to enter the continuum band and thus lead to so-called scattering resonance~\cite{PhysRevA.78.023617}.
%
% Revised on 2018/06/20
Although several equilibrium properties in hybrid atom-molecule systems have been studied, the QWs in these systems have not been revealed yet.
In particular, it is intriguing to explore the signature of DBS's via QWs.
%

% Introuduce our work.
In this article, by considering a one-dimensional (1D) Bose-Hubbard model with atom-molecular coupling, we study the QWs from two interacting Bose atoms occupying the same lattice site.
We focus on exploring the interplay among atom-molecule coupling, atom-atom interaction and atom-molecule energy detuning.
Without the atom-atom interaction, there are two kinds of DBS's supported by pure atom-molecule coupling.
Such an atom-molecule coupling may play the role of atom-atom interaction and then result in the correlated QWs.
Due to the atom-molecule energy difference, the atom-atom interaction can be balanced under certain resonant conditions and so that the DBS's are broken into scattering states.
Under strong interactions, the QWs show two light-cones corresponding to the two DBS bands.
By using the many-body degenerate perturbation theory, we give the effective models for the QWs of DBS's, in which the effective tunneling strengths of DBS's can be tuned by the atom-molecule energy difference.
Specifically, the interplay between tunnelings of atoms and molecule can suppress the nearest-neighbor (NN) tunneling of DBS's.

The paper is organized as follows.
In Sec.~\ref{two_systems}, we introduce our hybrid atom-molecule system and solve its energy bands.
In Sec.~\ref{hybridwalks}, we present the QWs from two atoms occupying the same site.
In particular, we discuss how the QWs are affected by the pure atom-molecule coupling~(\ref{couplingwalks}) and the interplay between atom-atom interaction and atom-molecule coupling~(\ref{intercouplewalks}).
In Sec.~\ref{dressedmodel}, we derive effective models for the QWs of DBS's and discuss the effective tunneling of DBS's.
At last, we make a brief summary and discussion of our results.

% ======= Model Section====
\section{HYBRID ATOM-MOLECULE ENERGY BANDS} \label{two_systems}
%%%%%%%%%%%

We consider two interacting Bose atoms in 1D optical lattices, where the two atoms can be converted into a molecular state via atom-molecule coupling.
The system obeys the Hamiltonian,
%======== Equation =========
\begin{eqnarray}
\hat H= &&\mathop -\sum \limits_{l=-L}^{L} \left( {{J_a}\hat a_l^\dag {{\hat a}_{l + 1}} + {J_m}\hat m_l^\dag {{\hat m}_{l + 1}} + H.c.} \right) \nonumber \\
&& + \frac{U}{2}\mathop \sum \limits_{l=-L}^{L} \hat n_l^a\left( {\hat n_l^a - 1} \right)+g\mathop \sum \limits_{l=-L}^{L} \left( {\hat a_l^\dag \hat a_l^\dag {{\hat m}_l} + H.c.} \right) \nonumber \\
&& +\mathop \sum \limits_{l=-L}^{L} \left( {{\varepsilon _a}\hat n_l^a + {\varepsilon _m}\hat n_l^m} \right).
\label{Hamiltonian}
\end{eqnarray}
%======== Equation =========
Here, $g$ is on-site atom-molecule coupling strength, $U$ is on-site background atom-atom interaction, $J_a(J_m)$ is the atomic(molecular) tunneling strength, $\varepsilon_a$($\varepsilon_m$) is the atomic(molecular) on-site energy, the lattice site index $l$ ranges from $-L$ to $L$, the total number of lattice sites is $L_t=2L+1$ and the periodic boundary condition (PBC) is imposed.
The bosonic operators $\hat a_l^\dag$ ($\hat m_l^\dag$) and $\hat a_l$ ($\hat m_l$) create and annihilate an atom (molecule) on the $l$-th site, respectively.
Compared with the atomic tunneling strength $J_a$, the molecular tunneling strength $J_m$ is much smaller and so that it can be neglected~\cite{PhysRevA.83.031607, PRA.71.043604, PhysRevLett.114.195302}.
Thus we set $J_m=0$ in our numerical calculations, but still keep it in our analytical calculations.
The atom-molecule coupling $g$ can be realized by applying magnetoassociation \cite{RevModPhys.82.1225} or photoassociation \cite{RevModPhys.71.1,PhysRevLett.80.4402} technique.
The on-site energies $\varepsilon_{a,m}$ can be tuned by applying external magnetic field.

The hybrid atom-molecule Hilbert space can be spanned by a complete set of orthogonal basis,
%======== Equation =========
\begin{equation}
{\cal H^{(\text 2)}} = \left\{ {\left| {{l_1}{l_2}} \right\rangle_a \oplus \left| j\right\rangle_m  } \right\}.
\label{eigeneqn:0}
\end{equation}
%======== Equation =========
%
Here, $\left| j\right\rangle_m = \hat m_{j}^\dag \left| {\bf{0}} \right\rangle (-L \le j \le L)$ denotes the state of one molecule in the $j$-th lattice site,
while $\left| {{l_1}{l_2}} \right\rangle_a= (1 + {\delta _{{l_1}{l_2}}})^{-1/2} \hat a_{{l_1}}^\dag \hat a_{{l_2}}^\dag \left| {\bf{0}} \right\rangle$ ($-L\le l_1\le l_2 \le L$) denotes the state of one atom in the $l_1$-th site and one atom in the $l_2$-th site, where $\delta_{l_1l_2}$ is Kronecker delta function.
Hence, one can expand the eigenstates as, $|\Phi \rangle  = {\sum _{{l_1'}\le {l_2'}}}{\phi _{{l_1'l_2'}}}|{l_1'}{l_2'}\rangle_a  + {\sum _{j'}}{\varphi _{j'}}|j'\rangle_m $.
Thus, the eigenstate problem $\hat H |\Phi\rangle=E|\Phi\rangle$ is described by the coupled equations
%======== Equation =========
\begin{eqnarray}
&&\sum _{l_1'\le l_2'}\phi _{l_1'l_2'} {_a}\langle {l_1l_2}|\hat H|{l_1'}{l_2'}\rangle_a + \sum\limits_{j'} {{\varphi _j}{}_a{{\langle {l_1}{l_2}|\hat H|j'\rangle }_m}}  = E \phi_{l_1l_2},\nonumber \\
&&\sum _{j'}{\varphi _{j'}}{_m}\langle j|\hat H|j'\rangle_m  +\sum\limits_{l{'_1} \le l{'_2}} {{\phi _{{l_1}{l_2}}}{}_m{{\langle j|\hat H|l{'_1}l{'_2}\rangle }_a}} = E\varphi_{j}.
\end{eqnarray}
For simplicity, we define $\psi_{l_1'l_2'}={( {1 + {\delta _{{l_1}'}}{{_l}_{_2'}}})^{1/2}} \phi_{l_1'l_2'}$ and so that the normalization coefficient is eliminated.
After some algebraic calculation, using commutation relations of  bosonic operators, one can obtain
%======== Equation =========
\begin{subequations}
\begin{eqnarray}
E{\psi _{{l_1},{l_2}}} &=&  - {J_a}\left( {{\psi _{{l_1},{l_2} + 1}} + {\psi _{{l_1} + 1,{l_2}}} + {\psi _{{l_1} - 1,{l_2}}} + {\psi _{{l_1},{l_2} - 1}}} \right),  \nonumber\\
&+& {\delta _{{l_1},{l_2}}}U{\psi _{{l_1},{l_2}}} + 2{\varepsilon _a}{\psi _{{l_1},{l_2}}} + 2g{\delta _{{l_1},{j}}\delta _{{l_2},{j}}}{\varphi _j},
\label{eigeneqn:1}
\end{eqnarray}
\begin{equation}
E{\varphi _j} =  - {J_m}\left( {{\varphi _{j + 1}} + {\varphi _{j - 1}}} \right) + {\varepsilon _m}{\varphi _j} + g{\delta _{{l_1}{j}}\delta _{{l_2}{j}}}{\psi _{{l_1},{l_2}}}.
\label{eigeneqn:2}
\end{equation}
\label{eigeneqn:3}
\end{subequations}
%======== Equation =========
Obviously, Eq.~\eqref{eigeneqn:1} and Eq.~\eqref{eigeneqn:2} show the hybridization of atomic and molecular states.
To solve these equations, we adopt the ansatz
%======== Equation =========
\begin{eqnarray}
\psi _{{l_1},{l_2}} &=& C_a{e^{iK_aR_a}}\xi (r), \nonumber \\
{\varphi _j} &=& C_m{e^{iK_mR_m}}.
\label{ansatz}
\end{eqnarray}
%======== Equation =========
Here, $K_a$, $R_a=(l_1+l_2)/2$ and $r=l_2-l_1$ are respectively the center-of-mass (c.o.m.) quasi-momentum, c.o.m. position and relative position of atoms.
Correspondingly, $K_m$ and $R_m=j$ are the molecular quasi-momentum and position, respectively.
The coefficients $C_{a}$ and $C_{m}$ are the normalization constants.
The function $\xi(r)$ is independent of $K_a$ and $R_a$,
\begin{equation}
\xi (r) =  {{C_ + }{e^{ik|r|}} + {C_ - }{e^{ - ik|r|}}},
\label{ansatz:1}
\end{equation}
where $k$ can be real or complex and $C_{\pm}$ are unknown coefficients.
From the physical point of view, the states of atoms $\psi _{{l_1},{l_2}}$ can be expressed as Bloch-like function with independent c.o.m. part and relative motion part.

Before we go further, let us prove that $K_a=K_m=K$ for eigenstates.
When $l_1 = l_2 = j$ ($R_m=R_a=R$), combining Eq.~\eqref{eigeneqn:2} and Eqs.~\eqref{ansatz}, we have
\begin{equation}
\frac{E + 2{J_m}\cos \left( {{K_m}} \right) - \epsilon_m}{g}C_m {e^{i{(K_m-K_a)}R}} = \xi \left( 0 \right){C_a}.
\label{KmKa}
\end{equation}
Because Eq.~\eqref{KmKa} holds for all $R \in [-L,L]$, we have $K_m = K_a$.
For simplicity, we denote $K_m = K_a=K$ and restrict it in the first Brillouin zone from now on.
Since the PBC requires $\psi_{l_1,l_2+L_t}=\psi _{l_1+L_t,l_2}=\psi_{l_1,l_2}$ and $\varphi _{j+L_t}=\varphi _j$, the c.o.m. quasi-momentum obeys $K=2\pi n/L_t$ with $n=-L,-L+1, \ldots,L$.

From Eqs.~\eqref{eigeneqn:3} and \eqref{ansatz}, introducing $\tilde E = E - 2{\varepsilon _a}$ and $\Delta  = {\varepsilon _m} - 2{\varepsilon _a}$, one can obtain
%======== Equation =========
 \begin{equation}
\tilde{E}\xi (r) = J_a^K\left[ {\xi (r + 1) + \xi (r - 1)} \right]+ {\delta _{r,0}}U_\mathrm{eff}\xi (r),
\label{EnergyEqn:1}
 \end{equation}
 %======== Equation =========
where $U_\mathrm{eff}={U + 2{g^2}/({{\tilde{E} - \Delta-J_m^K}}})$ and $J_a^K =  - 2J_a\cos (K/2)$, $J_m^K=-2J_m\cos (K)$.
Obviously, the atom-molecule coupling contributes an additional energy-dependent term in the effective interaction $U_\mathrm{eff}$.
This indicates that the atom-molecule coupling $g$ may play the role of atom-atom interaction $U$ and therefore DBS's may appear even the atom-atom interaction is absent.

In the case of $\Delta  \to \infty $ or $U \to \infty$, Eq.~\eqref{EnergyEqn:1} can be approximated as
%======== Equation =========
 \begin{equation}
\tilde{E}\xi (r) = J_a^K\left[ {\xi (r + 1) + \xi (r - 1)} \right]
 + {\delta _{r,0}}U\xi (r), \label{EnergyEqn:U0}
 \end{equation}
 %======== Equation =========
which reduces to the case of no atom-molecule coupling~\cite{jopB.41.16.161002}.

%---------------Figure----------------
\begin{figure*}%[htp]
  \includegraphics[width = \textwidth ]{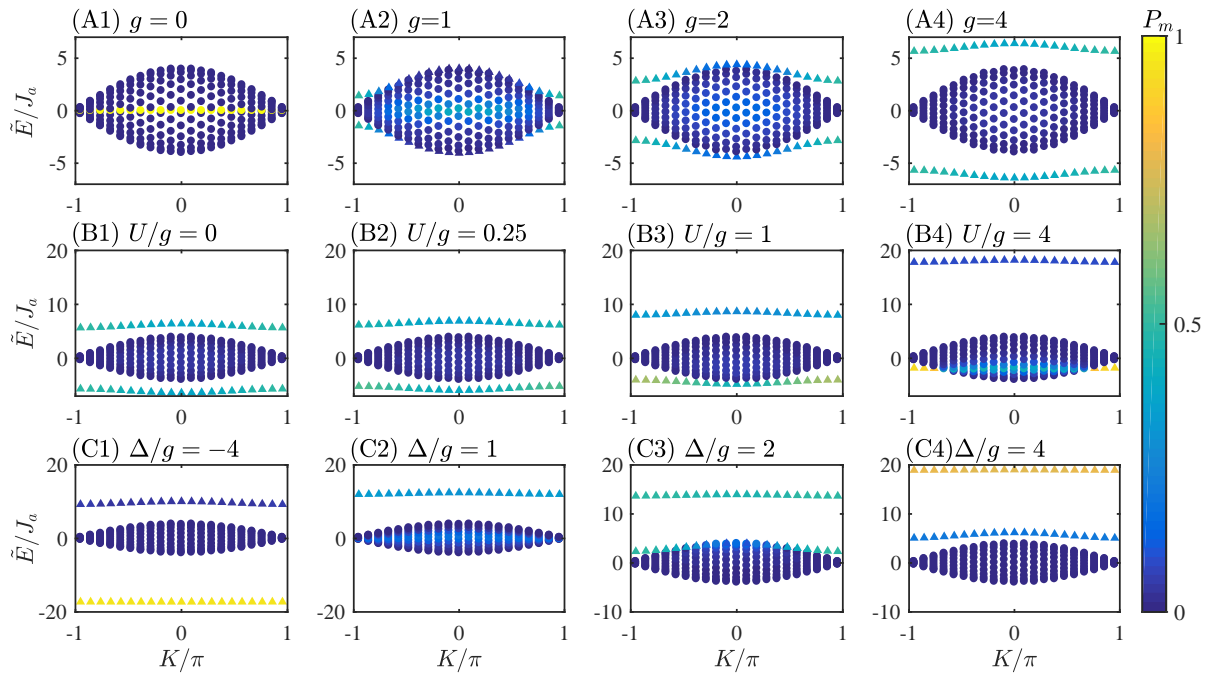}
  \caption{\label{fig:Spectrum}
  Energy spectrum under influence of atom-molecule coupling, atom-atom interaction and atom-molecule energy difference.
  The circular and triangular dots denote the scattering states and the DBS's, respectively.
  The color of each dot represents the proportion of molecular states, which is given by ${P_m} = \sum\nolimits_j {|{\varphi _j}{|^2}} $.
  (A1)-(A4): Energy spectrum for different atom-molecule coupling $g=0, 1, 2, 4$ with $U=\Delta=0$.
  (B1)-(B4): Energy spectrum for different atom-atom interaction $U/g=0, 0.25, 1, 4$ with $\Delta=0$ and $g=4$.
  (C1)-(C4): Energy spectrum for atom-molecule energy difference $\Delta/g=-4, 1, 2, 4$ with $g=4, U=8$.
  The other parameters are set as $J_a=1$, $J_m=0$ and $L_t=21$ by default.
  }
\end{figure*}
%---------------Figure----------------

\subsection{Continuum band}

The continuum band corresponds to scattering states whose $k$ are real numbers.
For a real $k$, substituting Eq.~\eqref{ansatz:1} into Eq.~\eqref{EnergyEqn:1}, we have the eigenenergies
%======== Equation =========
 \begin{equation}
\tilde E = 2J_a^K\cos (k).
\label{eigenergy:1}
\end{equation}
%======== Equation ========
%
Here, the value of $k$ can be determined by the following procedure.
Substituting Eqs.~\eqref{ansatz:1} and \eqref{eigenergy:1} into Eq.~\eqref{EnergyEqn:1},
one can find that the coefficients $C_{\pm}$ obey,
\begin{equation}
\frac{{{C_ + }}}{{{C_ - }}} =  - \frac{{ - {J_a^K}2i\sin k + \left( {U + \frac{{2{g^2}}}{{2{J_K}\cos k - \Delta  - J_m^K}}} \right)}}{{{J_a^K}2i\sin k + \left( {U + \frac{{2{g^2}}}{{2{J_K}\cos k - \Delta  - J_m^K}}} \right)}}.
\label{Cpm1}
\end{equation}
Furthermore, according to the PBC, $\xi(r)$ obeys $\xi(r+L_t)=e^{iKL_t/2}\xi(r)$ and therefore one can obtain the coefficients $C_{\pm}$
\begin{equation}
\frac{{{C_ + }}}{{{C_ - }}} =  - \frac{{{{\left( { - 1} \right)}^{iK{L_t}/2}} - {e^{ - ik{L_t}}}}}{{{{\left( { - 1} \right)}^{iK{L_t}/2}} - {e^{ik{L_t}}}}}.
\label{Cpm2}
\end{equation}
Combining Eqs.~\eqref{Cpm1} and \eqref{Cpm2}, one can determine $k$ by solving the following equation,
\begin{equation}
\frac{{  {J_a^K}2i\sin k - \left( {U + \frac{{2{g^2}}}{{2{J_K}\cos k - \Delta  - J_m^K}}} \right)}}{{{J_a^K}2i\sin k + \left( {U + \frac{{2{g^2}}}{{2{J_K}\cos k - \Delta  - J_m^K}}} \right)}} ={{( - 1)^{K{L_t}/2}}} {e^{-ik{L_t}}}.
 \label{k:analytical}
\end{equation}
Obviously, the above equation is invariant under the transformation $k \to - k$ and thus $k$ can be restrained in the region $[0,\pi]$.
Substituting the values of $k$ into Eq.~\eqref{eigenergy:1}, we obtain the eigenenergies of scattering states, which are denoted by the circular dots in Fig.~\ref{fig:Spectrum}.
%
% Added on 2018/06/08
%
Correspondingly, the explicit expression of $\xi(r)$ is given as
\begin{equation}
\xi (r) \sim {( - 1)^{K{L_t}/2}}{e^{ - ik{L_t}}}{e^{ik|r|}} + {e^{ - ik|r|}},
\end{equation}
which has the same form as the one of no atom-molecule coupling~\citep{PRA.90.062301,jopB.41.16.161002}.

Besides, we calculate the proportion of molecular state for each eigenstate,
\begin{equation}
  {P_m} = \sum\nolimits_j {|{\varphi _j}{|^2}},
\end{equation}
which is denoted by the color in Fig.~\ref{fig:Spectrum}. 
Due to the atom-molecule coupling, the scattering states are hybridization of molecular states and atomic states.

\subsection{Isolated bands}

Isolated bands correspond to the states with complex values of $k$.
If the atom-molecule coupling is absent, i.e. $g=0$, the atomic and molecular states are decoupled and there appears an isolated band corresponding to the molecular states, see Fig.~\ref{fig:Spectrum} (A1).
When $J_m=0$, the isolated molecular band is exactly given as $\tilde{E}=\Delta$.
For non-zero atom-molecule couplings $g$, the isolated bands correspond to DBS's, whose $k$ can be assumed as $k = \beta + i\eta$ (where $\beta$ and $\eta$ are both real numbers).
Noting that the wavefunction must remain finite when $r \to \infty $, Eq.~\eqref{ansatz:1} can be rewritten as
%======== Equation =========
\begin{equation}
\xi (r) = {e^{(i\beta  - \eta )|r|}}.
\end{equation}
For simplicity, we introduce ${e^{i\beta  - \eta }} \equiv {\alpha}$, which satisfies ${\alpha} \in \mathbb{C}$ and $0<|\alpha|<1$.
Thus $\xi(r)$ can be rewritten as
\begin{equation}
\xi (r) ={\alpha ^{|r|}}.
 \label{ansatz:2}
\end{equation}
This expression indicates that the wavefunctions of atomic states decay exponentially when the relative distance increases~\citep{jopB.41.16.161002}.
Combining Eqs.~\eqref{EnergyEqn:1} and \eqref{ansatz:2}, one can obtain
\begin{equation}
\tilde{E} = 2J_a^K\alpha  + \left( U + \frac{{2{g^2}}}{{\tilde E - \Delta -J_m^K}} \right)
\label{BoundBand1}
\end{equation}
for $r=0$, and
\begin{equation}
\tilde E = J_a^K({\alpha ^{ - 1}} + \alpha )
\label{BoundBand2}
\end{equation}
for $r>0$.
Here, $\tilde{E}$ and $\alpha$ are unknown parameters.
To ensure real eigenenergies $\tilde{E}$, the parameter $\alpha$ must be real as well and so that we have $\beta = m\pi$ and $m \in \mathbb{N}$.
By numerically solving Eqs.~\eqref{BoundBand1} and \eqref{BoundBand2}, we obtain two isolated bands for DBS's, see the triangular dots in Fig.~\ref{fig:Spectrum}.
The emergence of two isolated bands is consistent with the previous results obtained by other methods~\cite{PhysRevA.83.031607, PhysRevA.77.021601, PhysRevA.78.023617}.
From Eqs.~\eqref{BoundBand1} and \eqref{BoundBand2}, when $J_m=U=\Delta=0$, we find that if $(\tilde{E},\alpha)$ are their solutions, then $(-\tilde{E}, -\alpha)$ are also their solutions.
Furthermore, when the atom-molecule coupling strength $g$ increases, the two symmetric and isolated bands gradually separate from the continuum band, see Fig.~\ref{fig:Spectrum} (A1)-(A4), respectively.

\subsection{Interplay among the atom-molecule coupling, the atom-atom interaction and the atom-molecule energy difference} \label{interaction_energy}
\begin{figure}%[htp]
\includegraphics[width=\columnwidth]{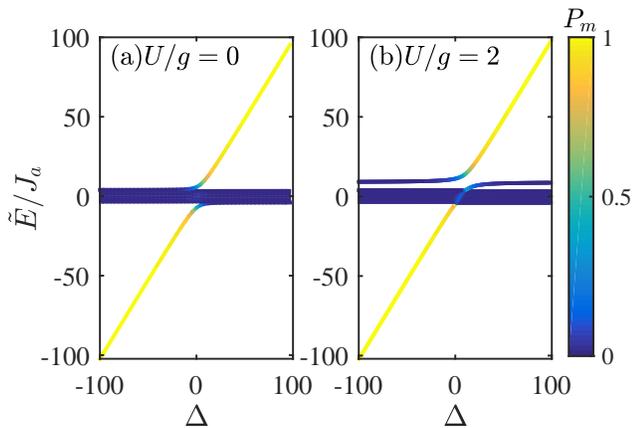}
  \caption{\label{fig:FixK0} Eigenenergies of the zero quasi-momentum states ($K=0$) versus the energy difference $\Delta$ and different ratios: (a) $U/g=0$ and (b) $U/g=2$.
  The color represents the proportion of molecular states, which is given by $P_m=\sum_j {|{\varphi _j}{|^2}}$.
  The other parameters are chosen as $J_a=1$, $\ J_m=0$, $g=4$ and $L_t=21$.
  }
\end{figure}

Below, given $g=4J_a=4$ and $J_m=0$, we will show how the atom-atom interaction ($U$) and the atom-molecule energy difference ($\Delta$) affect the energy spectrum.
To explore the interplay of $g$ and $U$, we choose $\Delta=0$.
For simplicity, we concentrate our discussion on the case of $U>0$.
Actually, the following discussion can be easily applied to the case of $U<0$.
We present the energy bands for different values of $U/g$ in Fig.~\ref{fig:Spectrum} (B1)-(B4).
Clearly, the repulsive interaction gradually lifts the energy of isolated bands.
Under strongly repulsive interaction, the lower isolated band enters into the continuum band and results in the resonance between scattering and bound states~\cite{PhysRevA.77.021601,PhysRevA.78.023617}, see  Fig.~\ref{fig:Spectrum}~(B4).
Around resonance, the states display strong hybridization than other states in continuum band.
When $U$ approaches to infinity, from Eq.~\ref{EnergyEqn:1}, the eigenenergies for the lower and upper isolated bands are given as $\tilde{E}=\Delta$ and $\tilde{E}=U$, respectively.
In this instance, the lower isolated band purely corresponds to the bare molecule, while the upper isolated band corresponds to the bounded atomic pair.
Given finite atom-atom interaction $U/g=2$, we then explore the interplay between $\Delta$ and $g$, as shown in Fig.~\ref{fig:Spectrum} (C1)-(C4).
When $\Delta/g \ll -1$, the upper and lower isolated bands are respectively dominated by the bounded atomic pairs and the molecular states, see Fig.~\ref{fig:Spectrum} (C1).
With the increase of $\Delta$, the lower isolated band is gradually shifted from the bottom to the upper of the continuum band, see Fig.~\ref{fig:Spectrum} (C2) and (C3).
Particularly, for certain values of $\Delta$, the lower isolated band may completely merge into the continuum band, as shown in Fig.~\ref{fig:Spectrum} (C2).
When $\Delta/g \gg 1$, the lower isolated band becomes dominated by the bounded atomic pair and the upper isolated band tends to be dominated by the molecular states, see Fig.~\ref{fig:Spectrum} (C4).
However, if the atom-atom interaction is zero, the lower isolated band will never merge into the continuum band.
To show this, we plot the eigenenergies for given c.o.m. quasi-momentum $K=0$ as a function $\Delta$, see Fig.~\ref{fig:FixK0}.
In the absence of atom-atom interaction ($U=0$), the lower (upper) isolated band gradually approaches to the bottom (above) boundary of the continuum band when $\Delta \rightarrow +\infty$ ($\Delta \rightarrow -\infty$), see Fig.~\ref{fig:FixK0} (a).
The two isolated bands for DBS's always sandwich the continuum band.
For non-zero atom-atom interaction ($U\ne0$), the energy of DBS's can merge into the continuum band, causing the resonance between DBS's and continuum band, see Fig.~\ref{fig:FixK0} (b).
In fact, one can prove that for a given $K$, there are two DBS solutions if $U=0$ and $g \ne 0$, while there can be only one solution if $U\ne0$, see Appendix.~\ref{apendix1} for more details.
To summarize, the atom-atom interaction is essential for the occurrence of the resonance.
%%

%-------------Figure---------------
%-------------Figure---------------

\subsection{Resonance between scattering states and DBS's} \label{resonance_condition}

In this subsection, we discuss the resonance between scattering states and DBS's and give the resonant conditions.
From Eqs.~\eqref{BoundBand1} and \eqref{BoundBand2}, one can give the energies for two isolated bands of DBS's.
However, as mentioned in the previous subsection, for non-zero atom-atom interaction $U$ we have proved that there may be only one solution under some specific conditions.
For a given $K$, the condition of only one solution of DBS's is given as
%
 %======== Equation =========
\begin{equation}
\frac{{2{g^2}}}{U} - 2|J_a^K| - J_m^K < \Delta  < \frac{{2{g^2}}}{U} + 2|J_a^K| - J_m^K.
\label{ineqn}
\end{equation}
%======== Equation =========
%
This indicates that there exists resonance between scattering states and DBS's.
If $J_m=0$, from Eq.~\eqref{ineqn}, one can find that there is only one DBS solution for all $K$ when $\Delta = 2g^2/U$, exactly corresponding to the result mentioned above in Fig.~\ref{fig:Spectrum} (C2).
This can be understood by the atom-molecule conversion in the limit of $J_a=J_m=0$, see Appendix.~\ref{atomic}.
By solving the eigen-equation, one can obtain three different kinds of eigenstates.
One kinds of the eigenstates corresponds to separated atomic states $|a_{l_1l_2}\rangle=|l_1l_2\rangle_a$ with $l_1 < l_2 $.
The other two kinds of eigenstates correspond to the \emph{dressed-molecule states}, which are in superposition of atomic state and molecular state ${|{d_l}\rangle  = A_{\sigma}|l{\rangle _m} + B_{\sigma}|l,l{\rangle _a}}$.
Here $A_{\sigma}$ and $B_{\sigma}$ are the coefficients of lower ($\sigma=1$) and upper ($\sigma=2$) dressed-molecule states.
The lower dressed-molecule states and the separated atomic states are degenerate when $\Delta  = 2g^2/U$ ($U>0$).
Under this condition, a tiny atomic tunneling will immediately make the separated atomic states into the atomic scattering states, and then the atomic scattering states couple with the dressed-molecule states.
That is why the degenerate condition is identical to condition where the lower isolated band merges into the continuum band.

\section{Hybrid atom-molecule quantum walks}\label{hybridwalks}

In this section, we analyze the QWs in our atom-molecule Hubbard system~\eqref{Hamiltonian}.
The initial state is chosen as $|\Psi(0)\rangle=|0,0\rangle_a$, in which both two atoms occupy the $0$-th lattice site.
The time-evolution is governed by the Schr{\"o}dinger equation,
\begin{equation}
  |\Psi(t)\rangle=e^{-i\hat H t}|\Psi(0)\rangle.
\end{equation}
The atomic and molecular density distributions are respectively defined as
\begin{eqnarray}
  n_{a,l}(t)&=&\langle \Psi(t)|a_l^{\dag}a_l|\Psi(t)\rangle, \nonumber \\
  n_{m,l}(t)&=&\langle \Psi(t)|m_l^{\dag}m_l|\Psi(t)\rangle.
\end{eqnarray}
The spatial correlation of atoms is characterized by a second-order correlation function,
%======== Equation =========
\begin{equation}
{\Gamma _{{l_1}{l_2}}}(t) = \langle \Psi (t)|{a^\dag _{l_1}}{a^\dag _{l_2}}{a_{{l_2}}}{a_{{l_1}}}|\Psi (t)\rangle,
\label{cor_fun}
\end{equation}
which relates to the probability $P_{l_1,l_2}(t)=|\langle l_1,l_2|\Psi(t)\rangle|^2$ via ${\Gamma _{{l_1}{l_2}}}(t)=(1+\delta_{l_1,l_2})P_{l_1,l_2}(t)$.
Thus ${\Gamma _{{l_1}{l_2}}}(t)$ gives the probability
of detecting one particle at $l_1$-th site and the other particle
at $l_2$-th site in the meantime.
The diagonal terms $\Gamma_{l_1=l_2}(t)$ describes the correlated QWs of two atoms, in which the two atoms walk as a whole.
The non-diagonal terms $\Gamma_{l_1\ne l_2}(t)$ describes the independent QWs of two atoms.
%%

%\subsection{QWs without atom-molecule coupling} \label{interactionwalks}
%%
If there is no atom-molecule coupling, the time evolution from the initial state $|0,0\rangle_a$ will evolve only in the subspace of the atomic states.
Since the molecular subspace is not involved, the QWs of atoms is expected to only depend on $J_a/U$.
When the atom-atom interaction is weak, the initial state has large overlaps with the atomic scattering states and so that the time-evolution is dominated by independent QWs~\citep{PRA.90.062301}.
When the atom-atom interaction is strong, the two atoms in the same site will form stable bound state and so that the time-evolution is dominated by correlated QWs~\citep{jopB.41.16.161002, PhysRevA.83.031607, fukuhara2013microscopic, PRA.90.062301}.

Indeed, under strong interaction, two atoms do perform correlated QWs, that is, the correlation function is dominated by the diagonal terms which recovers the results in Ref.~\citep{PRA.90.062301}.

\subsection{QWs with atom-molecule coupling}\label{couplingwalks}
\begin{figure}%[htp]
  \includegraphics[width=\columnwidth]{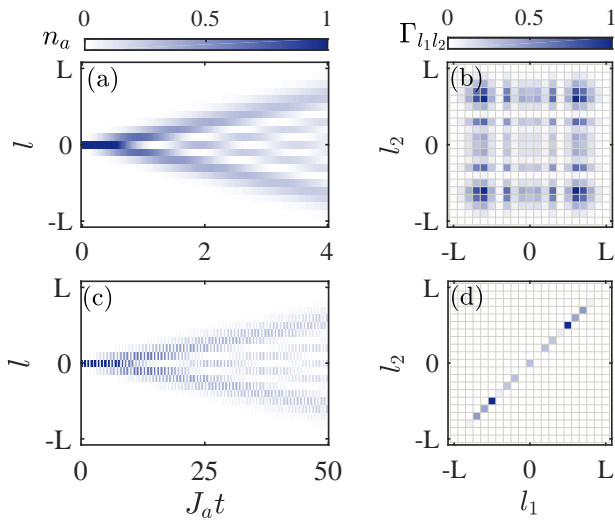}
  \caption{\label{fig:3rows}
  The QWs with: (a-b) zero atom-molecule coupling $g/J_a = 0$, and (c-d) strong atom-molecule coupling $g/J_a=10$.
  The left column shows the time-evolution of atomic density distribution and the right column show correlation functions of atoms for the final state.
  The other parameters are chosen as $J_a=1,J_m=0$, $U=0$, $\Delta=0$ and $L_t=21$.}
\end{figure}

Since the atom-molecule coupling may play the role of effective interaction, to show how the atom-molecule coupling affects the QWs, we turn off the atom-atom interaction ($U=0$) and the atom-molecule energy difference ($\Delta=0$).
For comparison, we simulate the QWs with $g=0$ and $g=10$.
The tunneling of atoms and molecule are chosen as $J_a=1,J_m=0$.
Without atom-molecule coupling, the time-evolution of atomic density distribution and the final correlation function are shown in Fig.~\ref{fig:3rows} (a) and (b).
The correlation function is dominated by the off-diagonal terms, which indicates that the two atoms walk independently.
At the presence of atom-molecule coupling, there will be the atom-molecule Rabi oscillations~\citep{PhysRevLett.99.033201,donley2002atom}.
If the atom-molecule coupling is strong enough, the atoms would go through many times of conversion before they walk to nearby lattice sites and thus experience a larger effective interaction.
In Fig.~\ref{fig:3rows} (c) and (d), we show the atomic density distribution and the final correlation function for $g=10$ and $\Delta=0$.
There appears notable stripes in the time-propagation of atomic density distribution, which can be explained by the fast atom-molecule conversion induced by strong atom-molecule coupling, see Fig.~\ref{fig:3rows} (c).
The strongly correlated QWs are also identified by the final correlation functions which are dominated by diagonal terms, see Fig.~\ref{fig:3rows} (d).
This is because the effective interaction is much larger than the tunneling strength, ${U_\mathrm{eff}} = 2{g^2}/( {\tilde E - \Delta  - J_m^K})\gg J_a$.

However, even for strong atom-molecule coupling, correlated QWs disappear when the atom-molecule energy difference $\Delta$ is much larger than the atom-molecule coupling $g$.
In such a situation, the larger atom-molecule energy difference makes the atom-molecule conversion negligible.
Therefore, atomic and molecular states are nearly decoupled and the two atoms walks independently since there is negligible effective atom-atom interaction from the atom-molecule conversion.

\subsection{QWs near the resonance between scattering states and DBS's} \label{intercouplewalks}

\begin{figure}%[htp]
  \includegraphics[width=\columnwidth]{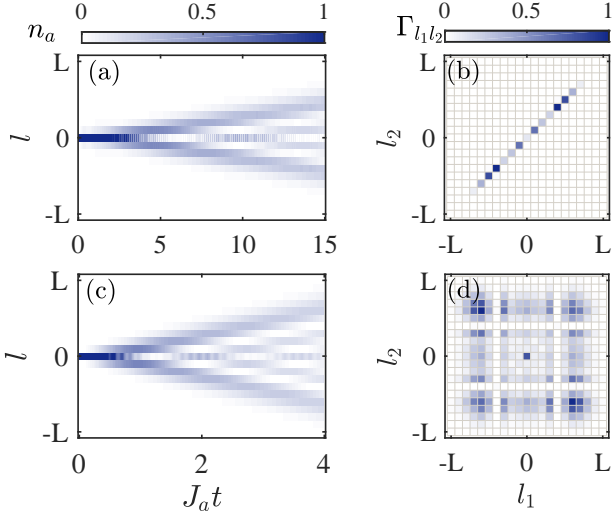}
  \caption{\label{fig:resonance}
  The hybrid atom-molecule QWs under: (a-b) non-resonant condition $\Delta = -40 \ll 2{g^2}/U$, and (c-d) resonant condition $\Delta= 40 = 2g^2/U$.
  The left column shows the time-evolution of atomic density distribution and the right column show correlation functions of atoms for the final state.
  The other parameters are chosen as $J_a=1$, $J_m=0$, $U=5$, $g=10$ and $L_t=21$.
  }
\end{figure}

In above, we show that the time-evolution are either dominated by independent QWs or correlated ones.
We wonder whether independent and correlated QWs may coexist.
As mentioned in Sec.~\ref{resonance_condition}, under the conditions of $g \gg J_{a,m}$ and $U \gg J_{a,m}$, the resonance between scattering states and DBS's takes place around $\Delta \simeq 2g^2/U$.
Below we will show the coexistence of independent and correlated QWs near the resonance between scattering states and DBS's.
Given $J_a=1$, $J_m=0$, $g=10$, and $U=5$, we present the QWs in non-resonant ($\Delta = -40 \ll 2{g^2}/U$) and resonant ($\Delta= 40 = 2g^2/U$) conditions, see Fig.~\ref{fig:resonance}.
Compared with  Fig.~\ref{fig:3rows} (c), there is no clear stripes in the time-propagation of atomic density distribution for large $\Delta$, see Fig.~\ref{fig:resonance} (a) and (c).
This is because large atom-molecule energy difference suppresses the atom-molecule conversion.
In the non-resonant condition, the diagonal elements of correlation function dominate after the time-evolution, indicating the strong co-walking behavior, see Fig.~\ref{fig:resonance} (b).
In the resonant condition, however, in addition to significant off-diagonal elements near the boundaries, there are significant diagonal elements on the diagonal line in the final correlation function, see Fig.~\ref{fig:resonance} (d).
This indicates the coexistence of independent and correlated QWs, although the propagation speed of correlated QWs is smaller than the one of independent QWs.
Such process can be explained by our argument in Sec.~\ref{resonance_condition}.
%%

%-------------Figure---------------
%-------------Figure---------------

%==== Perturbation Section ============
\section{EFFECTIVE SINGLE-PARTICLE MODEL FOR STRONGLY CORRELATED QUANTUM WALKS} \label{dressedmodel}

The strongly correlated QWs can be described by a single-particle model.
By employing the many-body quantum degenerate perturbation theory~\cite{JPC1977Takahashi}, we derive an effective single-particle Hamiltonian for the strongly correlated QWs.

To avoid the breakdown of DBS's near the resonance between scattering states and DBS's, we suppose $|\Delta  - 2{g^2}/U| \gg 0$.
When $J_{a,m} \ll g$ or $J_{a,m} \ll U$, the tunneling term $\hat T=-\sum \left( {{J_a}\hat a_l^\dag {{\hat a}_{l + 1}} + {J_m}\hat m_l^\dag {{\hat m}_{l + 1}} + H.c.} \right)$ in Hamiltonian ~\eqref{Hamiltonian} can be treated as a perturbation.
Defining the subspace $\mathcal{H}^d_{\sigma} =  \left\{ {|d_{\sigma,l}\rangle}, -L\le l \le {L}\right\}$ for DBS's (see Appendix.~\ref{atomic}), the projection operator is given by projecting the full Hilbert space $\mathcal{H}^{(2)}$ onto the unperturbed subspace $\mathcal H^d_{\sigma}$,
%======== Equation =========
\begin{equation}
{{\hat P}_{\sigma}} = \sum\limits_l {|{d_{\sigma,l}}\rangle \langle {d_{\sigma,l}}|} ,
\end{equation}
where $\sigma=\{1,2\}$ denotes the index for two different kinds of DBS's.
Besides, the projection onto the orthogonal complement of  $\mathcal H^d_\sigma$ reads as
\begin{eqnarray}
{\hat S_\sigma } &&= \sum\limits_{E_{{l_1}{l_2}}^{(0)} \ne E_\sigma ^{(0)}} {\frac{1}{{E_\sigma ^{(0)} - E_{{l_1}{l_2}}^{(0)}}}|{l_1}{l_2}\rangle \langle {l_1}{l_2}|}  \nonumber \\
&& + \sum\limits_{l,\sigma ' \ne \sigma } {{\frac{1}{E_\sigma ^{(0)} - E_{\sigma '}^{(0)}}}|{d_{\sigma ',l}}\rangle \langle {d_{\sigma ',l}}|} .
\end{eqnarray}
Therefore, according to the perturbation theory~\cite{JPC1977Takahashi} up to second order, we have
\begin{eqnarray}
{{\hat H}^{{\rm{eff}}}_\sigma} &&= {{\hat h}_{\sigma,0}} + {{\hat h}_{\sigma,1}} + {{\hat h}_{\sigma,2}} \nonumber \\
&&= {E_\sigma}{{\hat P}_\sigma} + {{\hat P}_\sigma}{{\hat T}}{{\hat P}_\sigma} + {{\hat P}_\sigma}{{\hat T}}\hat S_\sigma{{\hat T}}{{\hat P}_\sigma}.
\end{eqnarray}
%======== Equation =========
Substituting the projection operators and perturbation term into the above equation, we can obtain
%======== Equation =========
\begin{eqnarray}
{{\hat h}_{\sigma,0}} = &&{E_\sigma}\sum\limits_l {|{d_{\sigma,l}}\rangle \langle {d_{\sigma,l}}|}, \\
{{\hat h}_{\sigma,1}} =  &&- {J_m}{A_{\sigma}^2}\sum\limits_l {(|{d_{\sigma,l}}\rangle \langle {d_{\sigma,l + 1}}| + |{d_{\sigma,l + 1}}\rangle \langle {d_{\sigma,l}}|)},
\\
{{\hat h}_{\sigma ,2}} = &&\frac{{2{J_a}^2B_\sigma ^2}}{{E_\sigma ^{(0)} - E_{{l_1},{l_2}}^{(0)}}}\sum\limits_l {\left( {\begin{array}{*{20}{l}}
{2|{d_{\sigma ,l}}\rangle \langle {d_{\sigma ,l}}|+} \\
{  |{d_{\sigma ,l + 1}}\rangle \langle {d_{\sigma ,l }}| + h.c.}
\end{array}} \right)} \nonumber \\
 &&+ \frac{{{J_m}^2{A_1^2}{A_2^2}}}{{E_\sigma ^{(0)} - E_{\sigma '}^{(0)}}}\sum\limits_l {\left( {\begin{array}{*{20}{l}}
{2|{d_{\sigma ,l}}\rangle \langle {d_{\sigma ,l}}|+}\\
{  |{d_{\sigma ,l + 2}}\rangle \langle {d_{\sigma ,l}}| + h.c.}
\end{array}} \right)},
\nonumber
\\
\end{eqnarray}
%======== Equation =========
Here, the coefficients $A_{\sigma}$ and $B_{\sigma}$ are given by calculating the unperturbed time-independent Schr{\"o}dinger equation (see Appendix.~\ref{atomic}).

By introducing the mapping: $|{d_l}\rangle \langle {d_l}| \Leftrightarrow {d_l}^\dag {d_l},|{d_l}\rangle \langle {d_{l + 1}}| \Leftrightarrow {d_l}^\dag {d_{l + 1}},|{d_{l + 1}}\rangle \langle {d_l}| \Leftrightarrow {d_{l + 1}}^\dag {d_l}$,
the effective single-particle Hamiltonian can be written as
%======== Equation =========
\begin{eqnarray}
{{\hat H}^{{\rm{eff}}}_{\sigma}} &&=\sum\limits_l {\left({E_\sigma}+  { \frac{{4{J_a}^2{B_{1,2}^2}}}{{{E^{(0)}_\sigma} - {E^{(0)}_{l_1,l_2}}}}} +2\frac{{{J_m}^2A_1^2A_2^2}}{{E_\sigma ^{(0)} - E_{\sigma '}^{(0)}}}\right){d_{\sigma,l}}^\dag {d_{\sigma,l}}} \nonumber \\
&&  + \left( {\frac{{2{J_a}^2{B_{\sigma}^2}}}{E^{(0)}_\sigma-E^{(0)}_{l_1,l_2}} - {J_m}{A_{\sigma}^2}} \right)\sum\limits_j {\left( {{d_{\sigma,l}}^\dag {d_{\sigma,l + 1}} +H.c.} \right)} \nonumber \\
&& + \left( {\frac{{{J_m}^2A_1^2A_2^2}}{{E_\sigma ^{(0)} - E_{\sigma '}^{(0)}}}} \right)\sum\limits_l {\left( {{d_{\sigma ,l}}^\dag {d_{\sigma ,l + 2}} + H.c.} \right)}.
\label{ptb:effH}
\end{eqnarray}
%======== Equation =========
In addition to the nearest-neighbor (NN) tunneling, there appears the next-nearest-neighbor (NNN) tunneling, which originates from the effects of molecular tunneling.
The NNN tunneling brought by the atomic tunneling can be derived from 3rd-order perturbation theory, and we have neglected it since this term is extremely small compared with the lower order terms.
Since $|E_1^{(0)}-E_2^{(0)}| \gg |E_1^{(0)}-E_0^{(0)}|$ or $|E_2^{(0)}-E_0^{(0)}|$, the NNN tunneling term is generally negligible compared with other terms.
By implementing a Fourier transformation, the above single-particle Hamiltonian can be easily diagonalized and the eigenenergies are given as
%======== Equation =========
\begin{eqnarray}
{{E_{\sigma}^{{\rm{eff}}}}} =  && \left( {\frac{{8{J_a}^2{B_{\sigma}^2}}}{{{E^{(0)}_\sigma -E^{(0)}_{l_1,l_2}}}} - 4{J_m}{A_{\sigma}^2}} \right){\cos ^2}\left( {\frac{K}{2}} \right) \nonumber \\
&& + 4{\frac{{{J_m}^2A_1^2A_2^2}}{{E_\sigma ^{(0)} - E_{\sigma '}^{(0)}}}}\cos^2K+{E_{\sigma}^{(0)}} + {J_m}{A_{\sigma}^2},
\label{ptb:energy}
\end{eqnarray}
which are well consistent with the ones from numerical diagonalization of the original Hamiltonian.

\begin{figure}%[htp]
  \includegraphics[width=\columnwidth]{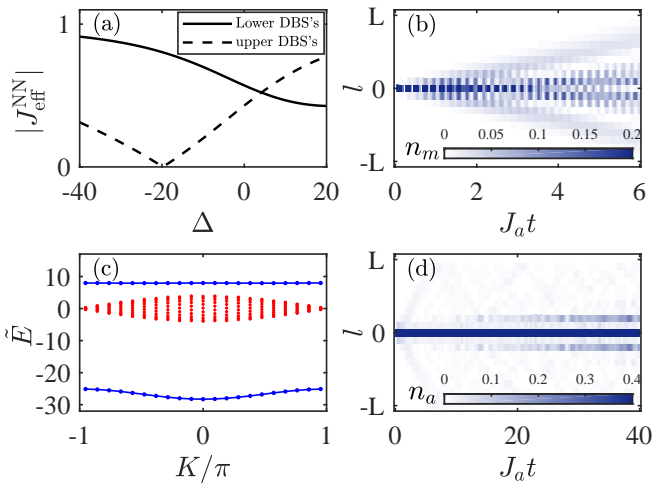}
  \caption{\label{fig:Jeff}
  (a) The effective nearest-neighbor tunneling strength $J^{NN}_{\rm{eff}}$ versus the atom-molecule energy difference $\Delta$.
  The parameters are chosen as $J_a=J_m=1$, $g=10$ and $U=0$.
  (b) Time-evolution of molecular density distribution with $\Delta=-10$ and the same parameters with (a).
  (c) The energy bands with $\Delta=-19.125$, $L_t=21$ and other parameters given in (a).
  The blue-dotted lines and the red dots correspond to the bands of DBS's and atomic scattering states respectively.
  (d) Long time-evolution of atomic density distribution with $\Delta=-19.125$ and other parameters given in (a).
  }
\end{figure}

In the effective single-particle Hamiltonian~\eqref{ptb:effH}, the effective NN tunneling strength is given as $J^{NN}_{{\rm{eff}},\sigma}= {{{2{J_a}^2{B_{\sigma}^2}}}/{(E^{(0)}_\sigma-E^{(0)}_{l_1,l_2})} - {J_m}{A_{\sigma}^2}} $.
Obviously, $J^{NN}_{{\rm{eff}},\sigma}$ also depends the atom-molecule energy difference $\Delta$.
In Fig.~\ref{fig:Jeff}~(a), we plot $J^{NN}_{{\rm{eff}},\sigma}$ as a function of $\Delta$, in which the solid and dashed lines respectively correspond to the upper and lower DBS bands.
The parameters are chosen as $J_a=J_m=1$, $g=10$ and $U=0$.
The effective tunneling strengths for the upper and lower DBS bands are always different except for the crossing point.
The different effective tunneling strength will result in different propagation speeds in QWs.
In Fig.~\ref{fig:Jeff} (b), we show the atomic density distribution with $\Delta=-10$ and other parameters as same as the ones for Fig.~\ref{fig:Jeff} (a).
Since the initial state mostly occupies the two DBS bands,
there appear two light-cones: the inner light cone and the outer one respectively correspond to the QWs of DBS's in the upper and lower bands.

%-------------Figure---------------
%-------------Figure---------------

%%
From Fig.~\ref{fig:Jeff} (a), near $\Delta = -19.125$, the effective tunneling strength of the DBS's in the upper band is almost zero, i.e. $J^{NN}_{\rm{eff}} \approx 0$.
Given $\Delta = -19.125$, we plot the energy bands in Fig.~\ref{fig:Jeff} (c).
The upper DBS's band is very flat, which indicates very small tunneling strength, while the lower DBS's band is not.
This is concordant with the results of effective model in Fig.~\ref{fig:Jeff}(a).
Noticing that $J^{NN}_{\rm{eff}} \approx 0$, there is only the NNN tunneling term ($J^{NNN}_{\rm{eff}} \simeq 0.005$) in the effective Hamiltonian~\eqref{ptb:effH}.
Therefore, the odd sites are never occupied in the QWs from the 0-th site, which is a clear significant of the NNN tunneling, see Fig.~\ref{fig:Jeff} (d).
Such novel phenomenon can be understood as the coherent interference between the atomic and molecular tunneling.
As shown in the perturbative calculation, the effective NN tunneling of DBS's can be achieved via two paths, one of which is the second-order atomic tunneling and the other one is the first-order molecular tunneling.
These two paths give rise to different values of effective tunneling energy.
When these two values have opposite values with the same magnitude, the total effective tunneling is cancelled out.
On the other hand, in the effective Hamiltonian~\eqref{ptb:effH}, the effective tunneling induced by the molecular tunneling is of first order, while the effective tunneling induced by atomic tunneling is of second order.
This means that, as the molecular tunneling may has considerable effects, it should be treated carefully in realistic systems.

% =========Final Summary Section==========

\section{summary and discussions}

In summary, we study the energy bands and hybrid atom-molecule QWs of a 1D coupled atom-molecule Hubbard system.
We find that the atom-molecule coupling can play the role of effective atom-atom interaction.
Unlike the conventional bounded atomic pair, the cooperation of the atom-atom interaction and the atom-molecule coupling induces two kinds of DBS's, which are the dressed molecule states in superposition of bounded atomic pair and bare molecule.
Even if the atom-atom interaction is absent, one can observe correlated QWs induced by the atom-molecule coupling.
Tuning the parameters (the atom-molecule energy difference $\Delta$, the atom-atom interaction $U$ and the atom-molecule coupling $g$) to satisfy the resonant condition, one of the DBS's will enter the continuum band and break into atomic scattering states.
Thus, one can observe the coexistence of independent and correlated QWs near the resonance between scattering states and DBS's.
Away from the resonant condition, we employ many-body quantum degenerate perturbation theory to derive the effective single-particle Hamiltonian for the two DBS bands.
The nearest-neighbor tunneling strength in the effective single-particle model can be turned off by tuning the atom-molecule energy difference $\Delta$.
Due to the two DBS's have different effective tunneling strengths, the QWs show two light cones with different propagation speeds.
%
% Added on 2018/06/20
Moreover, we find that the NN tunneling of one of the DBS's can be suppressed to zero due to the interference between atomic tunneling and molecular tunneling.
In this condition, the NNN tunneling become dominated and can be observed from the distribution of atomic density during the time-evolution.
Our study not only provides a full description for the hybrid atom-molecule QWs with atom-molecule coupling, but also will shine some light on the two-photon QWs with spontaneous parametric down-conversion (SPDC)~\citep{PhysRevLett.108.023601, ANTONOSYAN201422, PhysRevX.4.031007}.
In such a waveguide array, the near-degenerate signal and idler photons correspond to two identical atoms, the pump photon acts as the molecule, and the SPDC play the role of the atom-molecule coupling.
The difference is that, in the waveguide array, the energy of  signal and idler photons always equal to the pump photon, and there is no interaction between photons if the Kerr effects are absent.
According to our study, the idler and signal photons may have effective on-site interaction induced by the SPDC even if there are no Kerr effects~\cite{PhysRevLett.113.173601}.
Furthermore, the idler and signal photons can form dressed bound states with the pump photon when the SPDC is sufficiently strong.
Therefore, there will appear two different kinds of dressed photonic bound states with different effective hopping strengths between waveguides.

\begin{acknowledgments}
This work was supported by the National Natural Science Foundation of China (NNSFC) under Grants No. 11574405.
\end{acknowledgments}

% ========Some Supplement=========
\appendix

\section{Graphical illustration for solutions of DBS's} \label{apendix1}

\begin{figure}%[htp]
  \includegraphics[width=\columnwidth]{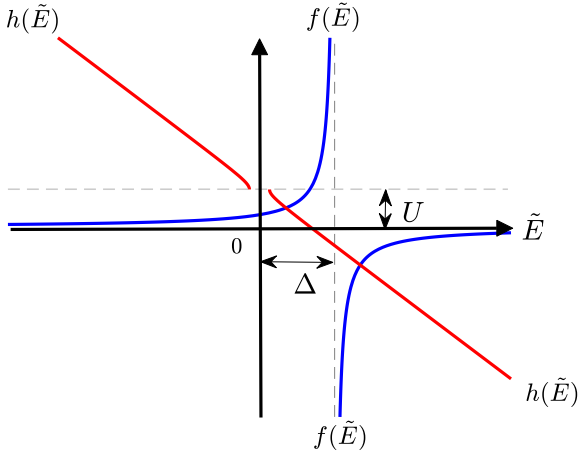}
  \caption{\label{fig:fE_appendix} The intersections of $f(\tilde{E})$ and $h(\tilde{E})$. Here, $g=10$, $U=20$, $J_a=1$, $J_m=0$ and $\Delta=50$.}
\end{figure}

To give the solutions of DBS's, one has to determine the parameter $\alpha$ by solving Eqs.~\eqref{BoundBand1} and \eqref{BoundBand2}.
From Eq.~\eqref{BoundBand2}, we have
\begin{equation}
\alpha^{\pm} (\tilde E) = \left({\tilde E \pm \sqrt {{{\tilde E}^2} - {{(2J_a^K)}^2}} }\right)/{2J_a^K},\label{A1}
\end{equation}
and $|\tilde{E}|>2J_a^K$.
Therefore, Eq.~\eqref{BoundBand1} can be rewritten as
%======== Equation =========
\begin{equation}
\tilde{E} - \frac{2{g^2}}{\tilde E - \Delta -J_m^K} = 2J_a^K\alpha^{\pm}(\tilde E)+U.
\label{eigenergy:3}
\end{equation}
%======== Equation =========
Because $|\alpha|<1$, we have $\alpha=\alpha^{-}$ when $\tilde E>0$ and $\alpha=\alpha^{+}$ when $\tilde E<0$.
Thus Eq.~\eqref{eigenergy:3} is equivalent to
%======== Equation =========
\begin{equation}
\frac{{ - 2{g^2}}}{{\tilde E - \Delta  - J_m^K}} =  \pm \sqrt {{{\tilde E}^2} - {{(2J_a^K)}^2}}  + U.
\label{eigenergy:4}
\end{equation}
Introducing
%========Equation=========
\begin{equation}
\left\{ \begin{array}{l}
f(\tilde E) \equiv  - 2{g^2}/(\tilde E-\Delta-J_m^K)\\
h(\tilde E) \equiv  \pm \sqrt {{{\tilde E}^2} - {{(2J_a^K)}^2}}  + U
\end{array} \right. ,
\label{functions_of_f_E}
\end{equation}
%========Equation=========
the solutions of Eq.~\eqref{eigenergy:4} can be obtained by solving $f(\tilde{E})=h(\tilde{E})$.
Therefore, the intersections of $f(\tilde{E})$ and $h(\tilde{E})$ give the solutions of $\tilde{E}$ and then the parameter $\alpha$ can be given from Eq.~\eqref{A1}.
In Fig.~\ref{fig:fE_appendix}, given $g=10$, $U=20$, $J_a=1$ , $J_m=0$ and $\Delta=50$, we show the intersections of $f(\tilde{E})$ and $h(\tilde{E})$.
Clearly, there are always two intersections if $U=0$, while there might be only one intersection if $U \ne 0$ for some values of $\Delta$.

%-------------Figure---------------
%-------------Figure---------------

\section{The frozen limit} \label{atomic}

In the frozen limit, $J_a=J_m=0$, the Hamiltonian reads
\begin{eqnarray}
\hat H &&= \frac{U}{2}\sum\limits_{l =  - L}^L {\hat n_l^a} \left( {\hat n_l^a - 1} \right) + g\sum\limits_{l = -L}^{{L}} {\left( {\hat a_l^\dag \hat a_l^\dag {{\hat m}_l} + H.c.} \right)} \nonumber \\
&&+ \sum\limits_{l =  - L}^L {\left( {{\varepsilon _a}\hat n_l^a + {\varepsilon _m}\hat n_l^m} \right)}
\label{atomic_Hamiltonian}
\end{eqnarray}
and there are two kinds of eigenstates
\begin{equation}
\left\{ \begin{array}{l}
\begin{array}{*{20}{c}}
{|{a_{{l_1},{l_2}}}\rangle  = |{l_1},{l_2}\rangle }&{1 \le {l_1} < {l_2} \le {L_t};}
\end{array}\\
\begin{array}{*{20}{c}}
{|{d_l}\rangle  = A|l{\rangle _m} + B|l,l{\rangle _a}}&{1 \le l \le {L_t}.}
\end{array}
\end{array} \right. ,
\end{equation}
where $A$ and $B$ are normalization coefficients.
By diagonalizing the Hamiltonian~\eqref{atomic_Hamiltonian}, one can obtain its eigenstates and eigenenergies.
The eigenenergy of $|{a_{{l_1},{l_2}}}\rangle$ is given as $\tilde{E}_0^{0}=0$.
For $|{d_l}\rangle$, we have
%======== Equation =========
\begin{subequations}
\begin{equation}
\left\{ \begin{array}{l}
\tilde E_1^{(0)} = \left( {U + \Delta  - \sqrt {8{g^2} + {{(U - \Delta )}^2}} } \right)/2 \\
C_1 = A_1/B_1= ({ \Delta - U - \sqrt {8{g^2} + {{(U - \Delta)}^2}}})/{2\sqrt 2 g} ,
\end{array} \right.
\label{H0:E1}
\end{equation}
\begin{equation}
\left\{ \begin{array}{l}
\tilde E_2^{(0)} = \left( {U + \Delta  + \sqrt {8{g^2} + {{(U - \Delta )}^2}} } \right)/2\\
C_2 = A_2/B_2 = ({\Delta - U  + \sqrt {8{g^2} + {{(U -\Delta)}^2}}})/{2\sqrt 2 g} ,
\end{array} \right.
\label{H0:E2}
\end{equation}
\label{atomic_limit}
\end{subequations}
%======== Equation =========
where $A_{\sigma} =  \pm \frac{C_{\sigma}}{{\sqrt {1 + {C_{\sigma}^2}} }}$, $B_{\sigma} =  \pm \frac{1}{{\sqrt {1 + {C_{\sigma}^2}} }}$ with $\sigma=\{1,2\}$.
Thus, the two eigenstates read as $|d_{\sigma,i}\rangle=A_\sigma|i{\rangle _m} + B_\sigma|i,i{\rangle _a}$.
These two states correspond to two isolated DBS bands.

\bibliography{Main}
\end{document}